\documentclass{article}

\usepackage{arxiv}

\usepackage[utf8]{inputenc} 
\usepackage[T1]{fontenc}    
\usepackage{hyperref}       
\usepackage{url}            
\usepackage{booktabs}       
\usepackage{amsfonts}       
\usepackage{nicefrac}       
\usepackage{microtype}      
\usepackage{cleveref}       
\usepackage{graphicx}
\usepackage{longtable,tabularx}
\usepackage{pdflscape}
\usepackage{natbib}
\usepackage{doi}
\usepackage{authblk}

\hypersetup{
    pdftitle={Seamless Digital Engineering: A Grand Challenge Driven by Needs},
    pdfsubject={eess},
    pdfauthor={James S.~Wheaton, Daniel R.~Herber},
    pdfkeywords={digital engineering, model-based systems engineering, systems engineering, grand challenge}
}

\title{Seamless Digital Engineering: A Grand Challenge Driven by Needs}

\author[1]{James S. Wheaton}
\author[1]{Daniel R. Herber}
\affil[1]{Department of Systems Engineering\\
Colorado State University\\
Fort Collins, Colorado, USA}
\date{}

\renewcommand{\headeright}{Preprint}
\renewcommand{\undertitle}{Preprint}

\begin{document}

\maketitle

\begin{abstract}
Digital Engineering currently relies on costly and often bespoke integration of disparate software products to assemble the authoritative source of truth of the system-of-interest. Tools not originally designed to work together become an acknowledged system-of-systems, with their own separate feature roadmaps, deprecation, and support timelines. The resulting brittleness and conglomeration of disparate interfaces in the Digital Engineering Ecosystem of an organization drains resources and impairs efficiency and efficacy. If Model-Based Systems Engineering were applied to this problem, a complete system architecture model would be defined, and a purpose-built computing system-of-systems would be constructed to satisfy stakeholder needs. We have decades of research in computer science, cybersecurity, software and systems engineering, and human-computer interaction from which to draw that informs the design of a Seamless Digital Engineering tooling system, but it would require starting from a clean slate while carefully adopting existing standards. In this paper, this problem space and solution space are characterized, defining and identifying Seamless Digital Engineering as a grand challenge in Digital Engineering research.
\end{abstract}

\section{Introduction}
Engineers and other staff and contractors primarily use \textit{tools} (see Table~\ref{tab:Definitions}) to produce engineered systems such as integrated circuits, spacecraft, and software that meet stakeholder needs. Increasingly, those tools are computer-based and therefore dependent on the wide range of commercially- and freely-available computer software and hardware. Digital Engineering (DE), in particular, is an effort to complete this transition and rely solely on a digitized Authoritative Source of Truth (ASoT) of the system-of-interest (SoI) for engineering and associated activities throughout the system development life cycle. However, due to the heterogeneity and other complexity factors of those computer-aided engineering (CAE) and ancillary tools, the achievement of DE is frustrated by integration and usability challenges.

\begin{table}[hp]
\caption{Definitions of related engineering terms (\textbf{emphasis} added).\label{tab:Definitions}}
\centering
\begin{tabularx}{\textwidth}{>{\hsize=.25\hsize}X X}
\textbf{Term} & \textbf{Definition} \\
\hline
Tool & ``The external employment of an unattached or manipulable attached environmental object \textbf{to alter more efficiently} the form, position, or condition of another object, another organism, or the user itself, when \textbf{the user holds and directly manipulates the tool} during or prior to use \textbf{and is responsible for the proper and effective orientation} of the tool.'' \cite{shumaker2011animal} \\
\hline
Technology & ``Artifacts made through a systematic application of knowledge and used to reach practical goals'' \cite{salomon1984technology} \textbf{``in a specifiable and reproducible way''} \cite{skolnikoff1993} \\
\hline
Engineering & ``The creative application of scientific principles to design or develop structures, machines, apparatus, or manufacturing processes, or works utilizing them singly or in combination; or to construct or operate the same \textbf{with full cognizance of their design}; or \textbf{to forecast their behavior under specific operating conditions}; all as respects an intended function, economics of operation \textbf{and safety to life and property}.'' \cite{engineers1947canons} \\
\hline
Systems \mbox{Engineering} (SE) & ``Interdisciplinary approach governing \textbf{the total technical and managerial effort} required to transform a set of customer needs, expectations, and constraints into a solution and to support that solution \textbf{throughout its life}.'' \cite{ISO24765} \\
& ``A transdisciplinary and integrative approach to enable the successful realization, use, and retirement of engineered systems,
using systems principles and concepts, and scientific, technological,
and management methods.'' \cite{INCOSEFellowsBriefing2019} \\
\hline
Model-Based \mbox{Systems Engineering} (MBSE) & ``The formalized application of modeling to support system requirements, design, analysis, verification and validation activities beginning in the conceptual design phase and continuing throughout development and later life cycle phases.'' \cite{incose2020vision} \\
\hline
Digital Engineering (DE) & ``An enterprise approach for incorporating data-driven infrastructure, methods and tools, innovation, and workforce transformation centered on an \textbf{authoritative source of truth (ASoT)} for engineering data.'' \cite{dod2018digital} \\
& ``DE is a transformation of the implementation of all engineering (and associated disciplines) across an enterprise to \textbf{take full advantage of the digital integration} of engineering work, data, knowledge, and wisdom across that enterprise.'' \cite{noguchi2020digital} \\
\hline
Authoritative Source of Truth (ASoT) & ``The authoritative source of truth captures the current state and the history of the technical baseline. It serves as the central reference point for models and data
across the lifecycle. The authoritative source of truth will provide traceability as the system of interest evolves, capturing historical knowledge, and connecting authoritative versions of the models and data.'' \cite{dod2018digital} \\
\hline
Digital Engineering Ecosystem (DEE) & The ``enterprises' interconnected digital environments, stakeholder-networks, and semantic data that allows the exchange of digital artifacts from an authoritative source of truth to serve the stakeholder communities' interests.'' \cite{DEE-DAU-Glossary} \\
\hline
Digital Engineering System (DES) & A theoretical engineered system-of-systems that provides the networked computing resources and human-computer interfaces for an enterprise to accomplish all digital engineering activities for successful realization of engineered systems. \\
\hline
\end{tabularx}
\end{table}

Many deficiencies in computing systems are tolerated because workarounds are available and then implemented, further entrenching their intractability. The intractability of DE is due to the high number of computing components and interfaces, the decentralized and inconsistent evolution of those components, and in particular, a profound lack of a cohesive system architecture model that systems engineers and other stakeholders would use to inspect, analyze, validate, verify, and modify the complex digital system. A useful way of thinking about these digital tooling deficiencies is with the concept of the \emph{reverse salient}, which may be defined as ``a set of critical problems'' \cite{hughes1993networks} or components that ``fail to deliver the necessary level of technological performance thereby inhibiting the performance delivery of the system as a whole'' \cite{dedehayir2008dynamics}, where the SoI is the Digital Engineering Ecosystem (DEE) that provides a set of affordances that depend on the underlying computing hardware and software components. The computing system has not been designed to meet the engineers' needs, but rather software components known as applications have been designed for general use. As systems engineers, we do not tolerate delivering defective systems that frustrate users with unpredictable behavior and random errors, so if we were contracted to develop our own Digital Engineering System (DES), we would naturally apply MBSE with full requirements traceability, and track the Quality Attributes to guarantee success.

Prior publicly-funded research initiatives have had similar goals. In the recent past, DARPA has funded such initiatives as CRASH (Clean-Slate Design of Resilient, Adaptive, Secure Hosts) \cite{DARPA_CRASH}, META-II \cite{DARPA_METAII} and CRAFT (Circuit Realization At Faster Timescales) \cite{DARPA_CRAFT}, recognizing that the systemic vulnerabilities and deficiencies of current computer systems and CAE tools require a clean-slate design to overcome. \cite{norman1999invisible} also recognized that ``the current paradigm is so thoroughly established that the only way to change is to start over again''. Given the advancements in Model-Based Systems Engineering (MBSE), computer science, integrated circuit design, and manufacturing over the past four decades, we have the opportunity to envision a fresh DES that meets our needs for safety, security, comprehensiveness, efficiency, and efficacy. The goal then ``is to move from the current situation of complexity and frustration to one where technology serves human needs invisibly, unobtrusively: the human-centered, customer-centered way'' \cite{norman1999invisible}, a goal that ought to resonate with systems engineers and computer users alike.

The primary contribution of this paper is to clarify the definition of \textit{Seamless Digital Engineering} in contrast with how DE is currently being implemented and identify it as a grand challenge in DE research. In Section \ref{sec:relatedWork}, historical and related research is summarized in relation to seamless DE. Section \ref{sec:problemSpace} characterizes the DE tooling problem space and current integration patterns for achieving DE outcomes. Section \ref{sec:definingSDE} defines seamless DE so it can be subject to further SE and DE research. Finally, Section \ref{sec:grandChallenge} discusses how the achievement of seamless DE would constitute a grand challenge in DE research typically beyond the capability of any one organization. Some research conclusions and a summary of future work follow.

\section{Related Work}
\label{sec:relatedWork}

The theoretical and practical foundations for DE have been developing for at least a half-century, dating back to ARPA-commissioned research by MIT on LISP machines \cite{lombardi1964lisp, bawden1977lisp, genera1987environment} which aimed to ``help designers get from prototype to product faster'' and whose ``key [was] an open architecture and highly-integrated development tools''. Along with Smalltalk computing systems \cite{ingalls1981design} that represented a similar trajectory in integrated tool-based computing environments, LISP machines are virtually nonexistent today, and we risk forgetting the many lessons and advancements made in those early years of computing. Later research has focused on \textit{ex post facto} integration of computer-based tools.

\cite{boehm1984software} described a UNIX-based integrated tooling system that assembled basic components provided by the OS in combination with custom software fit-to-task, not unlike what we practice today. \cite{thomas1992definitions} provides an essential formalism of tool integration that decomposes the concept into four: 1) process integration, 2) data integration, 3) presentation integration, and 4) control integration. That paper describes system properties, such as consistency, coherence, nonredundancy, and synchronization, that are still relevant today, and in particular, ``tools are well integrated when they have a common view of data''. Although these efforts focus on computer-aided software engineering (CASE), the principles and lessons are generalizable to DE.

Seamless model-driven systems engineering was first presented in Refs.~\cite{broy2009seamless,broy2010seamless} and later in \cite{Broy2020seamless}. \cite{broy2010seamless} identifies three ingredients to seamless integration, i.e., ``deep, coherent, and comprehensive integration of models and tools'': 1) a modeling theory that serves as the basis of the ASoT's semantic domain, 2) a complete architectural model that describes product and process, and 3) ``a manner to build tools that conform to the modeling theory and allow the authoring of the product model'', noting that the scientific basis is available but political barriers thus far prevent its realization. \cite{broy2010seamless} contributes critical theories and observations informed by automotive and aeronautical industry practice, such as a conceptual model for an integrated model engineering environment based on horizontal and vertical tooling aspects and the useful notion that currently available tools impose the current situation rather than the ideal case of formally-defined modeling theory dictating tool support. The rationale for this vision may be familiar to DE practitioners, such as: lack of model integration at both the conceptual and tooling level; transitions between artifacts are ad-hoc, unclear, and error-prone; redundancy of models that lose product information, leaving some implicit in engineers' minds; the high costs of internally developing tools and the unsatisfactory tool support for domain-specific modeling in the organization's core business; and that formal verification is essential to producing complex systems that are correct-by-design but that today we only benefit from ``islands of success''. \cite{broy2010seamless} uses overlapping views of \textit{seamless}: seamless model-based development, seamless systems engineering, and seamless system development, and offers the following definition:

\begin{quote}
``Seamless model-based development promises to lift software development to higher levels of abstraction by providing integrated chains of models covering all phases from requirements to system design and verification. In seamless model-based development, modeling is not just an implementation method but also a paradigm that provides support throughout the entire development and maintenance lifecycle.''
\end{quote}

This definition is software development-focused and lacks precision with respect to `higher levels of abstraction'; while it covers the complete system development lifecycle, it refers to ``integrated chains of models'' that somewhat contrasts with DE's ASoT; lastly, it highlights this approach as a \textit{paradigm} as opposed to solely an implementation method. \cite{Broy2020seamless} defines \textit{seamless} as an approach to software and system modeling where ``all models and all development steps are in a tight relationship with a clear understanding how they benefit'', but this definition appears to lack the precision needed to determine if the modeling approach is in fact seamless. Clarity of what qualifies as a ``tight relationship'' among models and steps would improve our understanding of seamless model-based development. Finally, the first published use of the term ``seamless digital engineering'' was by \cite{Perzylo2020SeamlessDE}, which describes knowledge integration as a key component to DE, but that nevertheless leaves the definition implicit and subject to the reader's interpretation of seamless.

\section{Characterizing the Problem Space}
\label{sec:problemSpace}

\begin{figure}[t]
\centering
\includegraphics[scale=0.85]{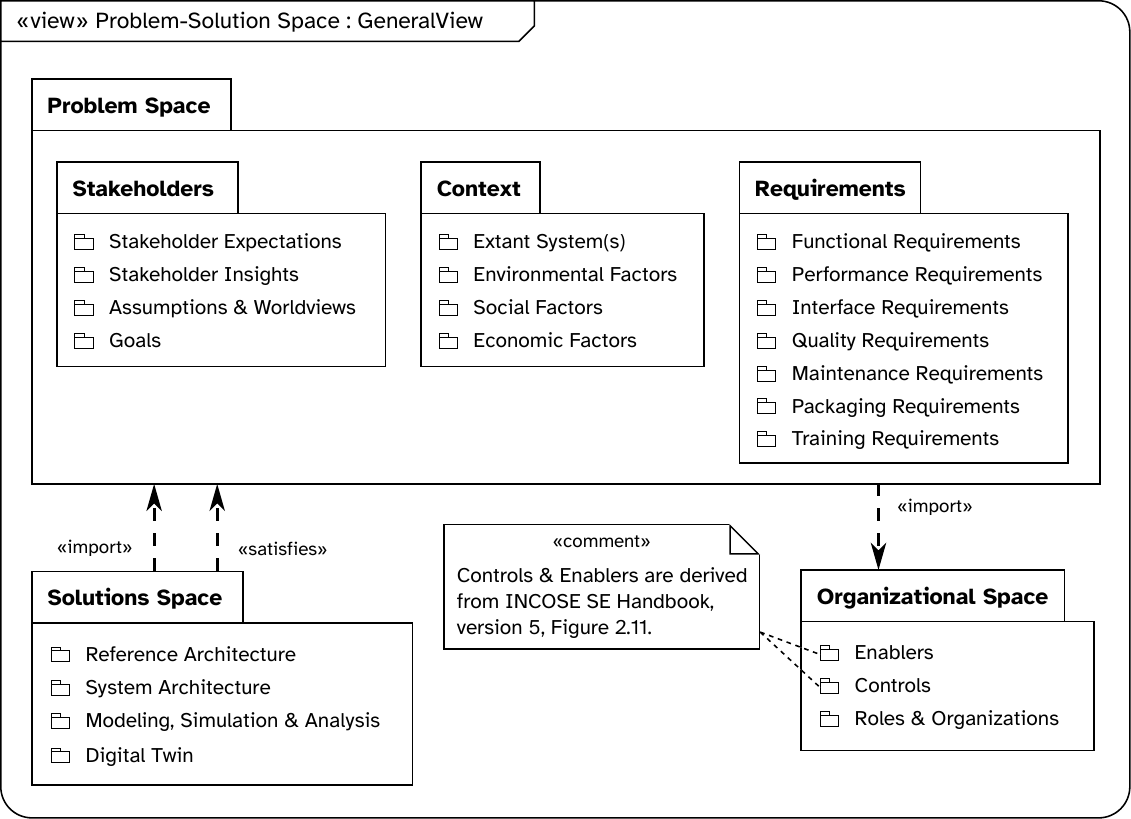}
\caption{Generic Problem-Solution Space (SysML v2 depiction).\label{fig:Problem-Solution_Space}}
\end{figure}

As \cite{norman1999invisible} succinctly notes, ``The personal computer is perhaps the most frustrating technology ever. The computer should be thought of as infrastructure''. Perhaps we do not think of computers as infrastructure but rather as suites of products. This mental model may be clouding our vision of future computing technology that suits our needs as engineers and communicators. Where much research after LISP machines has assumed the von Neumann computer architecture and widely-deployed operating systems as inviolable, a clean-slate approach favors the systems engineering practice of not making point-design decisions too early in the system development lifecycle. Let us first identify and characterize the problem space and then move on to proposing a solution space (see Fig.~\ref{fig:Problem-Solution_Space}).

\subsection{Stakeholders and Goals}
\label{sec:stakeholdersAndGoals}

\begin{table}[t]
\caption{Stakeholders of a DES.\label{tab:Stakeholders}}
\centering
\begin{tabularx}{\textwidth}{X X X}
  Engineer & Project/Program Manager & Accountant\\
  Acquirer & Contractor & Auditor\\
  Manufacturer & Technician & Designer\\
  Supplier & Regulator & Technical Writer\\
\end{tabularx}
\end{table}

Since DE encompasses the entire system development lifecycle, the stakeholders of a DES are varied and many. We provide here a list (see Table~\ref{tab:Stakeholders}) of potential stakeholders for a DES as a starting point, noting that some stakeholders listed, such as Engineer, may be further specialized depending on the role.

The original DE goals \cite{dod2018digital} are the visionary statements used to guide any DE implementation effort. The goals provide enough detail for us to link Stakeholder Expectations to them in a hierarchical, traceable relationship structure. Enumerated below, the DE goals are among the first elements to add to our system architecture model.

\begin{enumerate}
\item[\texttt{G1}] Formalize the development, integration, and use of models to inform enterprise and program decision-making.
\item[\texttt{G2}] Provide an enduring, authoritative source of truth (ASoT).
\item[\texttt{G3}] Incorporate technological innovation to improve the engineering practice.
\item[\texttt{G4}] Establish a supporting infrastructure and environments to perform activities, collaborate, and communicate across stakeholders.
\item[\texttt{G5}] Transform the culture and workforce to adopt and support DE across the lifecycle.\label{g5}
\end{enumerate}

While some aspects of the above DE goals depend on the organization's execution of the DE processes, e.g., ``transform the culture and workforce'', those goals may nevertheless be supported by a comprehensive DES that incorporates documentation and training resources.

\subsection{DE is an Adaptation of Existing Tools and Available Computing Systems}
\label{sec:DEasAdaptation}

Decades of competitive innovation in CAE tools have led to thousands of different software products in various functional areas throughout the product development life cycle. Startups entering high-tech fields such as integrated circuit design must first acquire a set of expensive software tools to begin their design work, with no guarantee that the various tools will be able to maintain a coherent system model. DE is an emerging paradigm where an ASoT of the SoI is maintained for the entire product development lifecycle, supporting Digital Thread and Digital Twin \cite{aiaa2020digital, aiaa2023digitalTwin}, and promising to streamline the engineering of complex systems. However, it prescribes no particular way of accomplishing this streamlined DE environment.

The burden of maintaining this organization-wide system of systems drains organization resources away from engineering systems: requiring data engineers and information technology administrators, more computing equipment, and the operational cost of vendor contracts that provide (and may eventually terminate) licenses and technical support. Moreover, staff is frequently switching between tools to accomplish their tasks, requiring productivity-draining context switches and training in each software product while being subject to vendors' upgrades that often modify or deprecate functions that have become integral to the organization's workflow.

\cite{karban2016creating, delp2019open} outline the OpenCAE tooling system, which amounts to an instantiation of the DE concept. Over 50 software products and languages are identifiable from the presentation in 2019. The presenter notes that the NASA Jet Propulsion Laboratory (JPL) uses ``14 server set-ups -- over 200 servers'' in cloud compute infrastructure and software-as-a-service to achieve their DEE. An acknowledged system-of-systems (ASoS) of this size and heterogeneity is likely to exhibit emergent fragility as the various software products are updated out-of-sync with one another. The academic and industry communities would benefit from more published research that studies the incongruities of computing system-of-systems of this kind. While the details of these deployed DEE are often proprietary, here we can acknowledge their complexity and the effort required to integrate the various components to achieve DE's goals.

\subsection{Common DE Integration Patterns}
\label{sec:integrationPatterns}

Software components can be made to work together typically using what is referred to as a `glue' or shim component. Even software that exposes the common UNIX-style plain text interface must nevertheless be integrated using a shim or glue component that parses the text, interprets any data, assembles a model, and then perform model transformations while striving to preserve model invariants though typically without any formal guarantees. The more humans are involved in these integration processes, the more human error is introduced, resulting in SoI defects that may compromise cost, schedule, and risk budgets.

\subsubsection{Import/Export Pattern}

\begin{figure}
\centering
\includegraphics[scale=0.85]{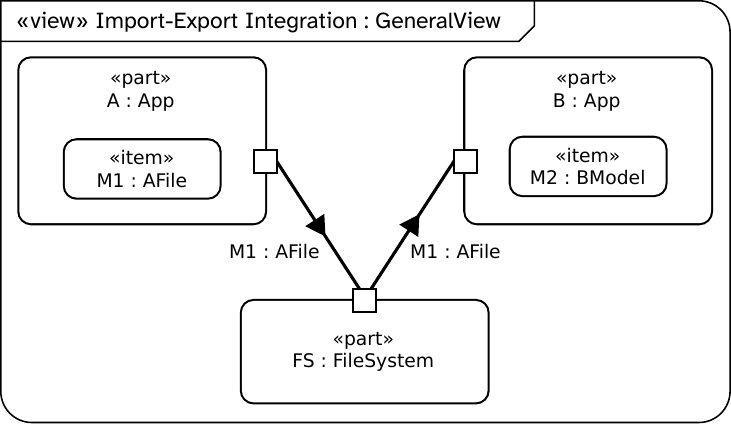}
\caption{Import/Export method of DE tool integration (SysML v2 depiction).\label{fig:importExport_method}}
\end{figure}

The most basic pattern of DE integration is import/export, where the engineering model is exported to a standardized or proprietary file format and then imported into another tool (see Fig.~\ref{fig:importExport_method}). This pattern relies on extra functionality in the tools, clear specification of the file format, and an accurate implementation thereof. Nevertheless, the pattern suffers when model information is lost during export and/or import due to incomplete implementations and incompatible meta-models.

\subsubsection{Custom Shim Pattern}

\begin{figure}
\centering
\includegraphics[scale=0.85]{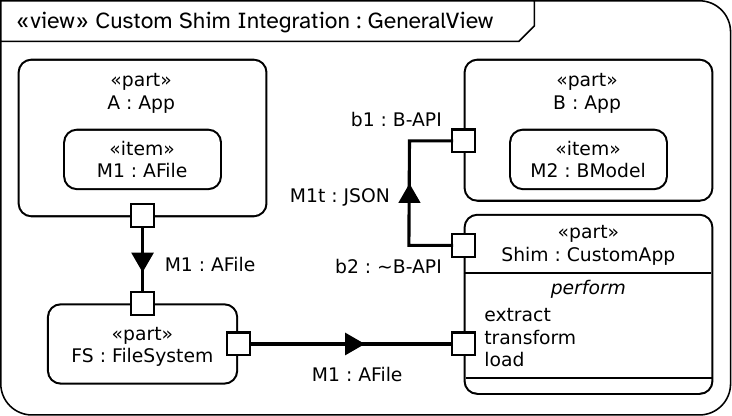}
\caption{Custom Shim method of DE tool integration (SysML v2 depiction).\label{fig:customShim_method}}
\end{figure}

When DE tools produce engineering artifacts with known schemas that are stored in known formats, custom shim software may be written to automate extract, transformation, load (ETL) operations (see Fig.~\ref{fig:customShim_method}). Examples of such artifacts include XML-based formats such as XMI and ReqIF, CSV or Excel spreadsheets, standard SQL databases, and JSON data that may loaded into a NoSQL database. The common drawback of this method, besides the additional software engineering effort, is that file formats and database schemas frequently change and often without due notice, resulting in brittle shims that must be urgently corrected to maintain the DE ASoS. Custom shims are also written when integrating software tools that expose RESTful or RPC-style APIs, with similar drawbacks when API specifications change.

\subsubsection{API Pattern}

\begin{figure}
\centering
\includegraphics[scale=0.85]{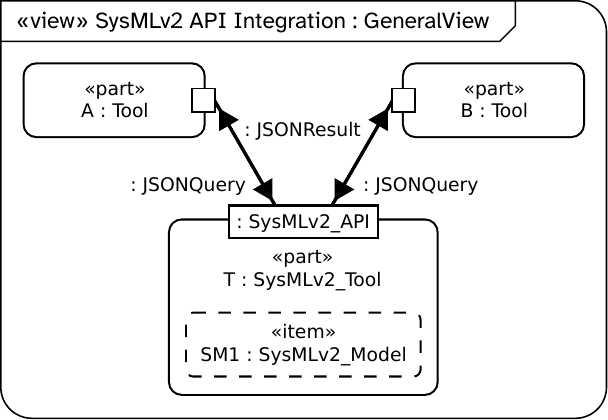}
\caption{SysML v2 API method of DE tool integration (SysML v2 depiction).\label{fig:API_method}}
\end{figure}

SysML v2 champions an API-based method of integration by exposing a RESTful interface (see Fig.~\ref{fig:API_method}) to the SysML model of the SoI \cite{bajaj2022systems}. Other DE tools are meant to use this RESTful API to query attributes and properties of the SoI model such that they are no longer duplicated in the separate tools, thereby reducing opportunities for emergent model definition divergence.

\subsection{DE As-Implemented Cannot be Elegant or Seamless\label{sec:DEnotSeamless}}

Before the commodification of computer hardware and software, computer-aided tools were designed to-purpose (SAGE \cite{EnticknapSAGE} is offered as an illustrative example).
Although the cost of such systems was extremely high due to custom development, they did not have as many computing standards and conventions to support and thus avoided the complexity of supporting and interfacing with non-essential or generalized technologies. Nowadays, it is often assumed that computing systems must support the plethora of computer technologies provisioned by Microsoft Windows or GNU/Linux to avoid `reinventing the wheel'.
This decision, while pragmatic, can suffer from a sunk-cost fallacy that limits the effectiveness of the deployed system and results in \textit{kludge}\footnote{Kludge is defined as ``a badly assembled collection of parts hastily assembled to serve some particular purpose (often used to refer to computing systems or software that has been badly put together)'' \cite{wordnet}.}.

Existing DE and CAE tooling systems lack elegance. \textit{System elegance} is a Quality Attribute (QAt) defined as ``a system that is robust in application, fully meeting specified and adumbrated intent, is well-structured, and is graceful in operation'' \cite{watson2020engineering, Watson2019Elegant, watson2014building, Madni2012Elegant}.
The four characteristics of elegant systems are \cite{griffin2010we}:

\begin{itemize}
\item System efficacy,
\item System efficiency,
\item System robustness, and
\item Minimizing unintended consequences.
\end{itemize}

DE as an ASoS lacks efficacy because the tools are nevertheless unaware of the ASoT and liable to introduce defects.
It lacks efficiency because the ASoS must itself be developed, deployed, and supported as a fixed and operational cost whose only opportunity for recoupment is the efficiency and quality gains not otherwise available.
Finally, its unintended consequences derive from the independent governance and development lifecycles of the adopted software components, whose emergent interactions as the SoS ages are impossible to predict with precision.

CAE tools, in general, do not consider the complete development lifecycle activities to be in-scope and only have plug-in and import/export support for other languages and tools. This scope limitation and typical reliance on proprietary data formats limit its efficacy and efficiency since the engineer must context-switch and adapt to the recalcitrant tool rather than the tool being adaptable to the engineer's needs. Additionally, import/export capabilities often have unintended consequences as even international standards may have ambiguity and often explicitly leave areas of the standard implementation-defined, so tools will have differing interpretations causing unintended consequences.

Generally speaking, seamless is defined as ``perfectly consistent and coherent'' and ``not having or joined by a seam or seams'' \cite{wordnet}, i.e., ``the fissure or gap formed by the imperfect union of two bodies laid or fastened together'' \cite{century_dictionary}. In systems engineering terms, the seam is the system architecture element that joins two or more interfaces and may employ parts and behaviors of its own to satisfy its traced requirements. So to be seamless, this joining element must not be necessary. This criterion has only limited satisfaction today, where multiple different model-based activities are possible from within the same tool (e.g., executable SysML and instrumented GUIs for system behavior simulations and early prototypes).

\subsection{Existing Computer Platforms are Nonviable}

\begin{figure}
\centering
\includegraphics[width=0.8\textwidth]{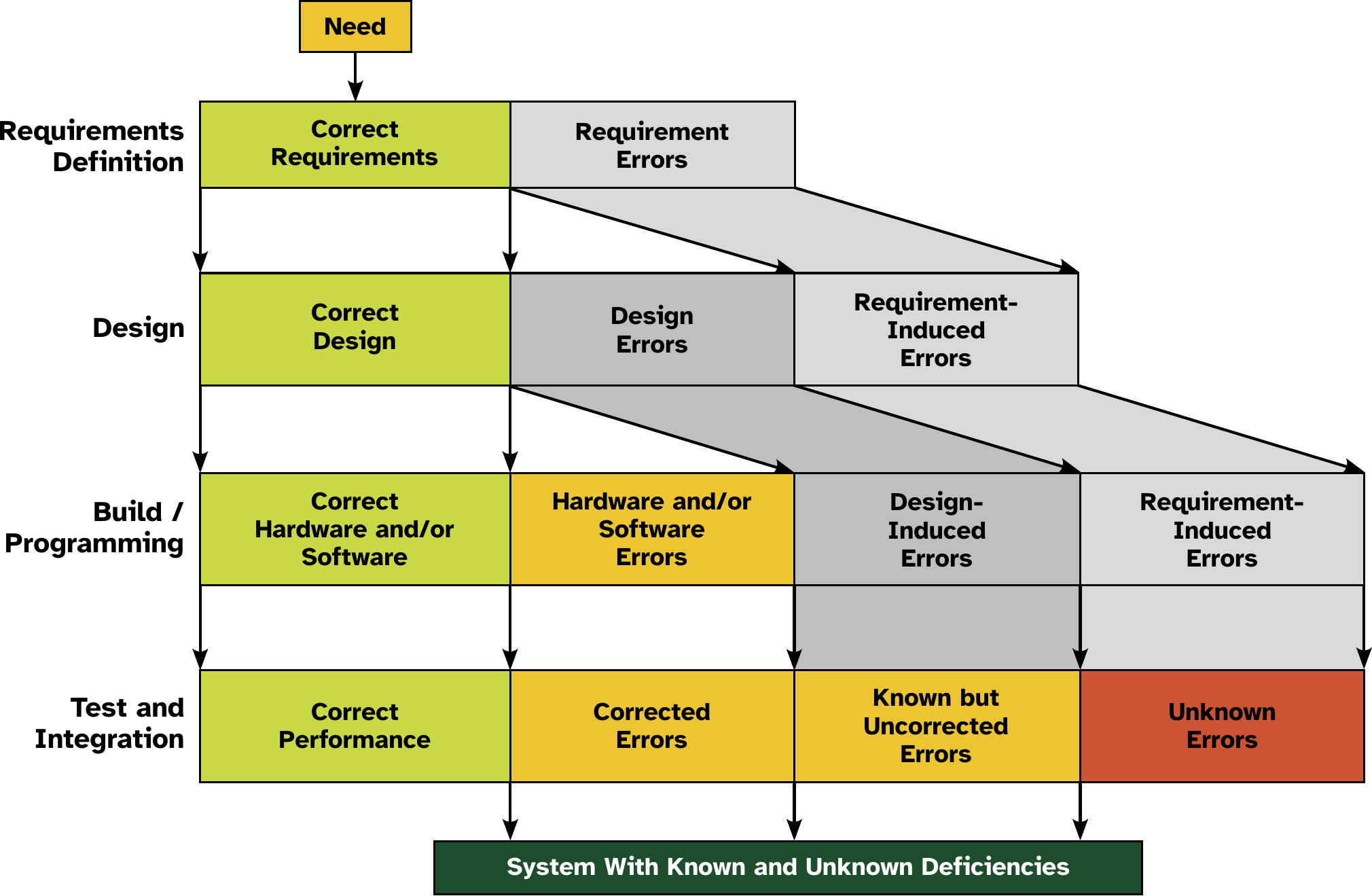}
\caption{The Error Avalanche (adapted from \cite{claxton2005test}).\label{fig:ErrorAvalanche}}
\end{figure}

A driving model for this research is The Error Avalanche as described in \cite{claxton2005test} and shown in Fig.~\ref{fig:ErrorAvalanche}: that not only will errors be introduced into the delivered system during the system development life cycle as a consequence of human error---even with the support of CAE tools---but the computer hardware and software used to design that system may also compromise the integrity of the system model and introduce subtle defects into the system. The existence of \textit{mercurial cores} that silently give the wrong calculation \cite{hochschild2021mercurial} illustrates this second point, a rather alarming discovery following the early Pentium FDIV bug \cite{halfhill1995error} and remotely-exploitable Meltdown \cite{lipp2020meltdown} and Spectre-class \cite{kocher2020spectre} CPU microarchitecture flaws that were predicted in \cite{porrasintel}.

By now it should be well-known that endpoint security is the weakest link in networked computer architecture and that the current approach to cybersecurity is nonviable \cite{ChiricescuSAFE, reubenstein2016semantically, von2015towards, snowden2019permanent, cisa_at_cmu}. And so uncritical adoption of existing computing technologies will eventually compromise enterprise security and thus compromise the robustness and `minimizing unintended consequences' properties of the DE tooling system. During a time when international competition is driving the adoption of MBSE and DE, and corporate espionage and data sabotage are serious risks in the presence of Advanced Persistent Threats (APTs), a clean-slate DE design must specify the full stack of computing hardware, firmware, and software to avoid making the same mistakes that we normally retain for backward-compatibility.

\section{Defining Seamless Digital Engineering}
\label{sec:definingSDE}

Prior descriptions of seamless DE or seamless MBE environments have left the contextual definition of seamless implicit \cite{broy2010seamless, Perzylo2020SeamlessDE}.
In this paper, the following definition of Seamless DE is offered with the additional criterion of end-to-end formal verification to guarantee seamless, error-free operation:

\begin{quote}
\textbf{Seamless Digital Engineering} (SDE) is a digital engineering tooling paradigm that guarantees model coherence and integrity by affording an elegant human-computer interface for systems modeling that is end-to-end formally verified down thru the computer hardware.
\end{quote}

\subsection{Seamless DE is a Refinement of DE}

In order to avoid the unmitigated complexity, hidden costs, and risks of existing computer systems, we must work to design a clean-slate DES that is capable of meeting stakeholder needs. There is a great opportunity here to avoid previous compromises and backward compatibility that have plagued existing computer systems, which are now the reverse salient that holds back the envisioned capability of DE and MBSE. Seamless digital engineering refines the definitions of DE quoted in Table~\ref{tab:Definitions} by adding the \textit{seamless} requirement for achieving the \textit{elegance} QAt as discussed above. We propose that this refinement may guide future research and collaboration toward a standardized and generational DES.

In light of the research summarized in Section \ref{sec:relatedWork}, a critical property of a successful DES must be that the ASoT or model integrity is guaranteed. A failure of integrity is an unintended change of the model data or ``the condition in which data are identically maintained during any operation, such as transfer, storage, and retrieval'' \cite{hapeman2001telecom}. Whether such a failure is corrected or not, it will inevitably add cost and risk to the system development project. The DES should be capable of detecting and automatically correcting errors wherever possible, leaving humans to their design activities rather than fixing and cleaning up after broken tools. The promise of computers in this respect has not yet been fully realized.

All the DE goals \cite{dod2018digital} are addressable by an adequate system architecture. A seamless DES would be designed with those goals defined in the system architecture model to support their bidirectional traceability for verification and validation. Although DE goal \hyperref[g5]{\texttt{G5}} regarding ``transform the culture'' is beyond the scope of the system, an adequate architecture model of the SDE system would include an operational viewpoint showcasing how common operational processes (workflows) would be accomplished using the system. And while DE metrics are now collated and processed with great effort, a capable SDE system would provide toggle-able instrumentation for process and product quality measurements with automated, configurable dashboards for monitoring DE effectiveness across the organization or project.

The vision of computers as invisible infrastructure \cite{kay1972dynabook, norman1999invisible} still seems worlds away when compared to our daily experience. And so, with respect to the presentation and process integration described in \cite{thomas1992definitions}, seamless integration must also encompass human-computer interface (HCI) ergonomics. We may even strive for computing systems that are \textit{delightful} to use and help guide and focus the user on their activities and tasks by incorporating HCI and human factors engineering research into the design.

\subsection{Seamless Integration Requirements}
\label{sec:seamlessRequirements}

Tool integration problems are central to DE transformation efforts at organizations, and they influence workforce attitudes toward change. Stakeholders become familiar with their tools---including the quirks---and how they fit together to satisfy their needs. So naturally, proposed changes to those workflows are often met with skepticism and resistance. An important goal of SDE is to smooth out those seams (defined in Section~\ref{sec:DEnotSeamless}) so stakeholders can focus on productive work rather than developing workarounds that are coping mechanisms for deficient tooling.

\begin{landscape}
{\footnotesize
\begin{table}
\caption{Transforming tool integration questions \cite{thomas1992definitions} into system requirements.\label{tab:transformedRequirements}}
\begin{tabularx}{\columnwidth}{ >{\hsize=.25\hsize}X >{\hsize=.28\hsize}X X X }
\textbf{Dimension} & \textbf{Aspect} & \textbf{Question} & \textbf{Transformed Requirement(s)} \\
\hline
Presentation \mbox{Integration}
& Appearance and behavior
& To what extent do two tools use similar screen appearance and interaction behavior?
& The SDE Tools shall conform to the [SDE HCI Standard].\\
& Interaction paradigm
& To what extent do two tools use similar metaphors and mental models?
& Following HCI Customization by the Operator, the SDE System shall apply the Operator HCI Configuration uniformly.\\
\hline
Control \mbox{Integration}
& Provision
& To what extent are a tool's services used by other tools in the environment?
& The SDE System shall provide common functionality as formally-verified System Libraries.\\
& Use
& To what extent does a tool use the services provided by other tools in the environment?
& The SDE Tools shall reuse System Libraries for Tool-specific functions.\\
& & & The SDE Tools shall provide Tool functions conforming to the [SDE Tool Interface Standard].\\
\hline
Process \mbox{Integration}
& Process step
& How well do relevant tools combine to support the performance of a process step?
& The SDE System shall provide the Workflow Definition Tool.\\
& Event
& How well do relevant tools agree on the events required to support a process?
& The Workflow Definition Tool shall define the Process Event using [SDE Data Definition Standard].\\
& Constraint
& How well do relevant tools cooperate to enforce a constraint?
& During Workflow Definition \& Execution, the Workflow Definition Tool shall enforce the Process Constraint on the Workflow-Defined Tool.\\
\hline
Data \mbox{Integration}
& Interoperability
& How much work must be done for a tool to manipulate data produced by another?
& The SDE Tools shall define the applicable Data Schema conforming to the [SDE Data Definition Standard].\\
& Nonredundancy
& How much data managed by a tool is duplicated in or can be derived from the data managed by the other?
& The SDE System shall provide the ACID-compliant System Database.\\
& Data \mbox{consistency}
& How well do two tools cooperate to maintain the semantic constraints on the data they manipulate?
& The SDE System Database shall maintain semantic constraints of data where available.\\
& \mbox{Data exchange}
& How much work must be done to make the nonpersistent data generated by one tool usable by the other?
& The SDE Tools shall write transactions to the SDE System Database for each change performed by the Operator.\\
& Synchronization
& How well does a tool communicate changes it makes to the values of nonpersistent, common data?
& The SDE System Database shall provide the Transaction Log applicable to the SDE Tool.\\
\hline
\end{tabularx}
\end{table}
}
\end{landscape}

The tool integration evaluation questions from \cite{thomas1992definitions} may be transformed into system requirements as in Table~\ref{tab:transformedRequirements}. These requirements serve to define the problem space (Fig.~\ref{fig:Problem-Solution_Space}) with respect to SDE tool integration. One aspect commonly discussed in the context of DEE is interoperability, and integration patterns such as a custom shim (Fig.~\ref{fig:customShim_method}) are often developed to transform data without having access to data schemas and definitions. The transformed requirement with respect to interoperability states that SDE Tools shall define the data schema used according to a standard such that other tools may parse it directly and avoid rework and mistakes. Likewise, the Control Integration dimension leads to the requirement that SDE Tools provide what is essentially a model-generated Interface Control Document for other tools to consume and use directly. This level of architecture openness and model-based interface documentation would help resolve many tool integration challenges.

As SysML v2 requirement definitions, the transformed requirements in Table~\ref{tab:transformedRequirements} may then be reused as requirement usages in various levels of the system architecture providing bidirectional traceability. SysML v2 requirements are modeled as constraints so that the system either satisfies the requirement or not. While the questions in Table~\ref{tab:transformedRequirements} often ask ``how well?'', in a clean-slate architecture, the components can be designed to satisfy these measures as requirements. The tool integration questions may also be modeled directly as SysML v2 \textit{concerns} that are similar to requirements but include one or more stakeholders and allow the relevant requirement(s) to \textit{frame} the concern. Further SysML v2 meta-modeling could capture the dimensions and aspects in Table~\ref{tab:transformedRequirements}, giving them ontological definitions and relations consistent with systems engineering practice.

\subsection{Seamless Integration Design Pattern}
\label{sec:seamlessDesignPattern}

\begin{figure}
\centering
\includegraphics[scale=0.85]{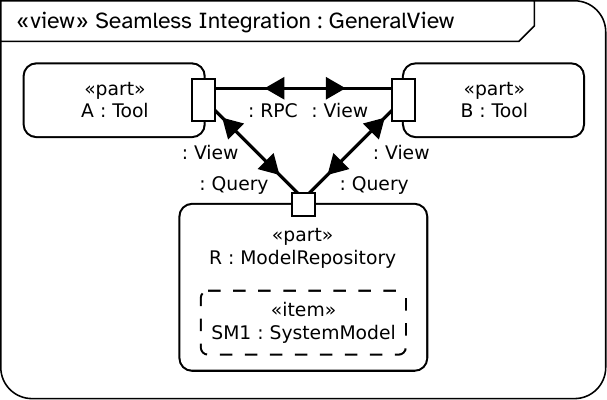}
\caption{Seamless design pattern of DE tool integration (SysML v2 depiction).\label{fig:seamless_method}}
\end{figure}

A seamless integration design pattern (see Fig.~\ref{fig:seamless_method}) uses system-wide standard interfaces, data schemas, and object representations in a live DES to accomplish the integration of SoI models to form the ASoT. By drastically reducing the number of interfaces, tools are designed to speak the same language. With self-documenting interfaces, tools may query other tools to discover and perform the available functions. When the ASoT is queried by a tool, a view is returned including the unique identifiers---or rather---handles of live objects. In this integration design pattern, it is important to note that data is not copied and divorced from its source that requires later merging and reconciliation.

\subsection{Seamless DE Architecture Tenets}
\label{sec:seamlessArchTenets}

\begin{figure}
\centering
\includegraphics[width=\textwidth]{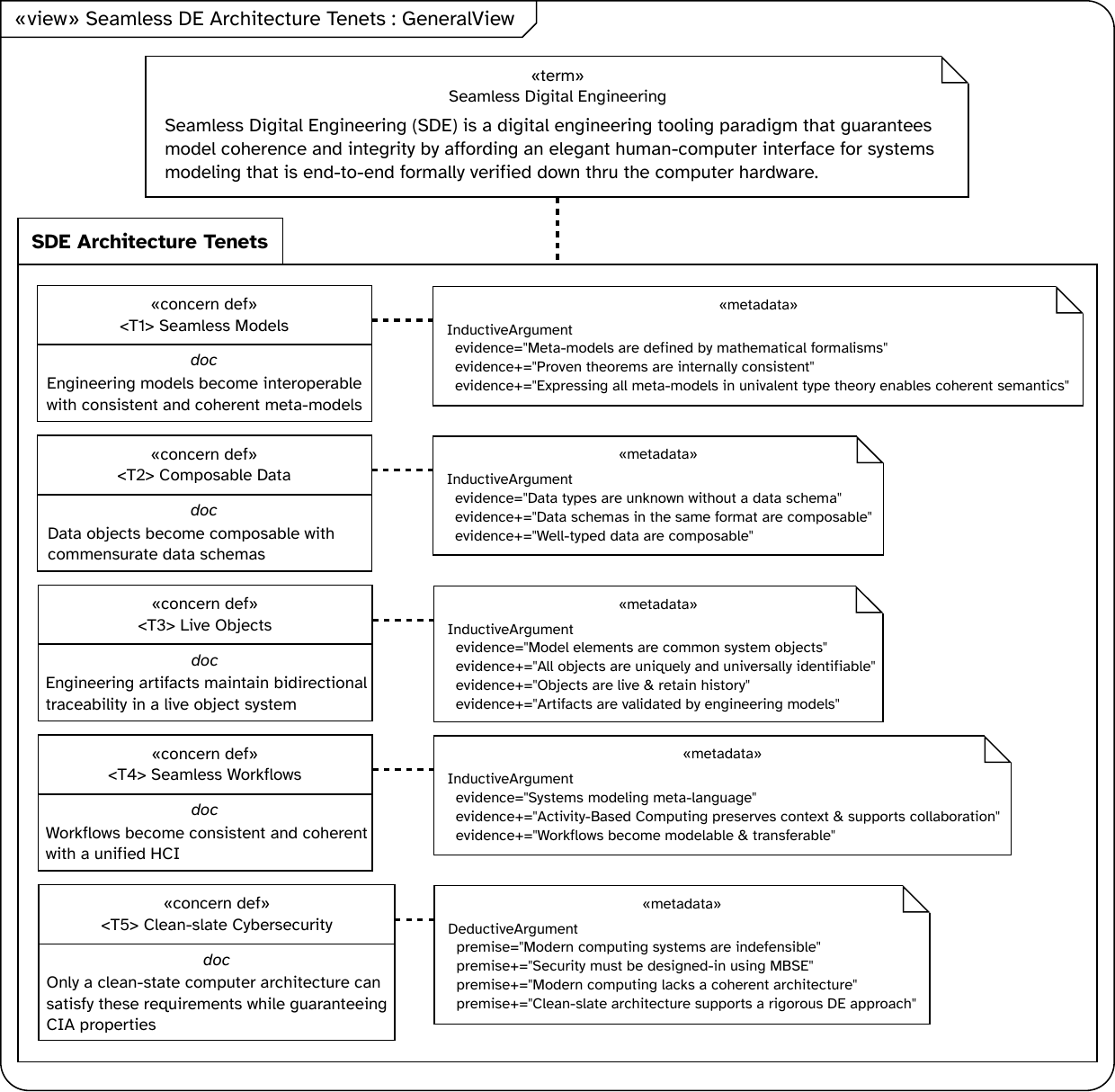}
\caption{Seamless DE Architecture Tenets as packaged concern definitions (SysML v2 depiction).\label{fig:SDE_Architecture_Tenets}}
\end{figure}

Seamless DE is based on a set of insights or architecture tenets that will guide its detailed architecture definition. These tenets are derived from observations of extant computing systems and existing research in computer science, human-computer interaction, and cybersecurity. Figure~\ref{fig:SDE_Architecture_Tenets} shows how these architecture tenets may be modeled as SysML v2 concerns with their own subjects, stakeholders, attributes, and constraints --- to be framed by system requirements in a reference architecture model. In summary, the architecture tenets that enable SDE are: 1) Seamless Models, 2) Composable Data, 3) Live Objects, 4) Seamless Workflows, and 5) Clean-slate Cybersecurity.

By packaging and presenting these insights together, they may be better understood than if analyzed individually. In accordance with systems engineering principles, capturing system architecture information as model data that is then presented in a variety of stakeholder views is superior to doing one or the other alone. Again, further SysML v2 meta-modeling may better capture, connect, and present the SDE architecture tenets --- our intent is to demonstrate how modeling these subjects can aid stakeholder understanding that improves system development outcomes. Rather than existing as implicit knowledge harbored by system designers, the SDE architecture tenets are explicit knowledge captured in the ASoT.

\section{A Grand Challenge Driven by Needs}
\label{sec:grandChallenge}

Grand challenges have historically been proposed to identify important research directions in a field and to help focus research and development efforts. \cite{hoare2003verifying} has clarified the criteria of what makes a research challenge a `grand challenge' in particular. Table~\ref{tab:grandchallenge} summarizes how SDE, as proposed by the authors of this paper, fits those criteria. By proposing SDE as a DE research direction using this framework, it may be better understood in the context of past grand challenges.

A grand challenge inspires collaboration across multiple fields and challenges researchers to adopt a transdisciplinary approach. Transdisciplinary systems engineering \cite{madni2019transdisciplinary} is especially well-suited for tackling the mess \cite{ackoff1997systems} of DE. Seamless DE draws on research in computer science, computer engineering, human-computer interaction, knowledge engineering, formal methods, and MBSE to envision a DES designed to meet stakeholder needs. Systems engineering researchers are encouraged to learn from these fields and apply that learning with systems praxis \cite{systemsPraxis} to the SDE grand challenge.

As a grand challenge, the work involved is beyond the capability of any one enterprise. The economics are challenging and would require investment from multiple organizations working in concert. The formal verification effort, in particular, has not been achieved at the proposed scale --- although we may rest assured that a sufficiently specified component that is formally verified is 'finished,' unlike the software we use from day-to-day, allowing engineers to progress steadfastly. Another salve is that with clean-slate design, workarounds of legacy systems are unnecessary, and we may dream beyond such reverse salients as the central processing unit (CPU) and POSIX standard interfaces \cite{posix1995}. Emulating open-source software projects may not be enough to manage the complexity, and new approaches to open-source systems engineering may need to be tested. Lastly, thinking on a generational time scale may alleviate the apprehension of accomplishing such a megaproject.

\begin{table}
\caption{Evaluation against criteria for a grand challenge \cite{hoare2003verifying}.\label{tab:grandchallenge}}
\begin{tabularx}{\textwidth}{ l X }
\textbf{Criterion} & \textbf{Rationale} \\
\hline
\textit{Fundamental} & Elegant and correct engineering tools are fundamental to development of complex systems. \\
\textit{Astonishing} & An all-encompassing DE tooling system does not exist, but it is at least conceivable and astonishing if realized. \\
\textit{Testable} & Systems engineering test \& evaluation methods apply, and the prototype system shall be continuously verified and validated. \\
\textit{Inspiring} & People want better tools because it makes their work easier. The scope is grand and inspiring. \\
\textit{Understandable} & Varieties of computer-aided tools are widely in use, and they may be understandably synthesized. \\
\textit{Useful} & The application of SE in this way shall inform computer science and related fields, as well as being eminently useful as a tool.  \\
\textit{Historical}  & This challenge is related to early research in computing contracted by the US DoD, and ties in with former grand challenges in computing.  \\
\textit{International}  & Everyone is welcome to contribute as everyone may potentially benefit from an open-source tool.  \\
\textit{Revolutionary}  & Systems, software and computer engineering could experience a paradigm shift as the approaches augment each other within a cohesive tooling system. \\
\textit{Research-directed}  & Advanced research is drawn upon and brought into an elegant system architecture in a scope commercially infeasible. \\
\textit{Challenging}  & While isolated tools and methods are identifiable, their synthesis will present new challenges and open new inquiries for research. \\
\textit{Feasible}  & Recent advancements in MBSE, programming languages and formal verification make a SDE system development feasible. \\
\textit{Incremental}  & By using MBSE to specify system components and behaviors, and the program itself, progress may be made incrementally and in parallel efforts. \\
\textit{Co-operative}  & Researchers are already working in the relevant fields, and may wish to contribute in this new context. \\
\textit{Competitive}  & Point-design decisions are unclear in some cases, such as the best type system or virtual machine, and competitive entries may be evaluated against the architecture metrics. \\
\textit{Effective} & Potentially it will change how we advance DE research. \\
\textit{Risk-managed}  & Even failures are valuable research outcomes. By using SE, we can reorient at every spiral. \\
\hline
\end{tabularx}
\end{table}

\section{Conclusions and Future Work}
\label{sec:conclusion}

Rather than coping with fragile, byzantine computing systems for implementing DE, we could have a DES by virtue of a computing system that is designed to satisfy user needs. Systems engineers are especially well-suited for tackling this design problem, but it will require a transdisciplinary systems engineering \cite{madni2019transdisciplinary} and an international collaborative approach. While many would contest a clean-slate architecture approach as infeasible, there are interacting emergent effects of modern networked computing systems that cannot be resolved with patches or by adding more components: namely, cybersecurity, reliability, and HCI considerations. By incorporating decades of research in computing and adjacent fields, we can design in the required functionality at the system level and eliminate unnecessary complexity that triggers The Error Avalanche (Fig.~\ref{fig:ErrorAvalanche}). A clean-slate design effort of this nature has been proposed in the past \cite{DARPA_CRASH, ChiricescuSAFE}, but it warrants renewed investment and research through the DE lens.

We contend that under the umbrella of ``Seamless Digital Engineering Grand Challenge'', the necessary resources can be assembled to design a DES that meets DE practitioners' needs. In this paper, we defined SDE as a paradigm (Section~\ref{sec:definingSDE}) to help guide future research, and we distinguished it from prior uses in the literature with the inclusion of the elegance QAt and end-to-end formal verification for guaranteeing ASoT coherence and integrity. Seamless integration requirements (Table~\ref{tab:transformedRequirements}) were derived from tool integration concerns in the literature to demonstrate how SE methods may address this problem. A seamless design pattern of DE tool integration (Fig.~\ref{fig:seamless_method}) was depicted in SysML v2 to contrast this paradigm with existing tool integration patterns (Figs.~\ref{fig:importExport_method}, \ref{fig:customShim_method}, and \ref{fig:API_method}) that cope with tooling seams. Finally, SDE architecture tenets (Fig.~\ref{fig:SDE_Architecture_Tenets}) were derived to provide a concise set of concerns that guide system design.

As a grand challenge, significant work lies ahead. Comprehensive research is needed in eliciting stakeholder needs, transforming those needs into a finite set of system requirements, developing a reference architecture with SysML v2, and constructing prototypes for early evaluation. SysML v2 has been chosen to position this work in line with the next-generation of MBSE tools. An open-source reference architecture model is being developed to invite further feedback and collaboration. In this way, DE researchers and practitioners from around the world can contribute to a seamless DES that inspires and gratifies.

\bibliographystyle{unsrtnat}
\bibliography{main}

\end{document}